\documentclass[a4paper,twoside,obeyspaces]{article}

\usepackage{url}
\usepackage{xcolor}
\usepackage{subfigure}
\usepackage{alltt}
\usepackage{fancyhdr}
\usepackage{graphicx}
\usepackage{wrapfig}
\usepackage{amstext}
\usepackage{amsmath}  % align env
\usepackage{amssymb}  % \mathbb
\usepackage{amsthm}
\usepackage{mathtools}% underbracket
\usepackage{multicol}
\usepackage{pslatex}
\usepackage{apalike}
\usepackage{stmaryrd} % \llbracket \rrbracket
\usepackage{newfloat} % DeclareFrameBox
\usepackage{framed}   % framebox
\usepackage{SCITEPRESS}     % Please add other packages that you may need BEFORE the SCITEPRESS.sty package.

\makeatletter
\@ifpackageloaded{txfonts}\@tempswafalse\@tempswatrue
\if@tempswa
  \DeclareFontFamily{U}{txsymbols}{}
  \DeclareFontFamily{U}{txAMSb}{}
  \DeclareSymbolFont{txsymbols}{OMS}{txsy}{m}{n}
  \SetSymbolFont{txsymbols}{bold}{OMS}{txsy}{bx}{n}
  \DeclareFontSubstitution{OMS}{txsy}{m}{n}
  \DeclareSymbolFont{txAMSb}{U}{txsyb}{m}{n}
  \SetSymbolFont{txAMSb}{bold}{U}{txsyb}{bx}{n}
  \DeclareFontSubstitution{U}{txsyb}{m}{n}
  \DeclareMathSymbol{\aleph}{\mathord}{txsymbols}{64}
  \DeclareMathSymbol{\beth}{\mathord}{txAMSb}{105}
  \DeclareMathSymbol{\gimel}{\mathord}{txAMSb}{106}
  \DeclareMathSymbol{\daleth}{\mathord}{txAMSb}{107}
\fi
\makeatother

\def\KPU{KPU}

\makeatletter
\newcommand{\customlabel}[2]{%
\protected@write \@auxout {}{\string \newlabel {#1}{{#2}{}}}}
\makeatother

\def\E{{\cal{E}}}
\def\enc[#1]{\E[#1]}
\def\D{{\cal{D}}}
\def\dec[#1]{\D[#1]}
\def\QED{\raisebox{0.8ex}{\framebox{\kern0.2ex}}}
\def\C#1{{\mathbb C}[#1]}
\def\S#1{{\mathbb C}^L[#1]}

\DeclareFontShape{OT1}{cmtt}{bx}{n}
     {
      <5> <6> <7> <8> <9>
      <10> <10.95> <12> <14.4> <17.28> <20.74> <24.88> cmbtt10
      }{}
\DeclareFontShape{OT1}{cmtt}{b}{n}
  {<->sub * cmtt/bx/n}{}

\makeatletter
\def\blfootnote{\xdef\@thefnmark{}\@footnotetext}
\makeatother

\DeclareFloatingEnvironment[fileext=box,placement={!tb},name=Box]{mybox}

\newenvironment{frameenv}[1][]
{%
 \begin{mybox}[#1]%

 \begin{framed}
 \begin{minipage}{0.99\columnwidth}

 \begin{footnotesize}
}{%
 \end{footnotesize}\end{minipage}\end{framed}\end{mybox}%
}

\begin{document}
\setlength\textfloatsep{2ex}

\title{An Obfuscating C Compiler for Encrypted Computing}

\author{\authorname{%
        {Peter T. Breuer}
}
\affiliation{{Hecusys LLC, Atlanta, GA}, USA}
%\email{ptb@hecusys.com}
}

\abstract{
This paper describes an `obfuscating' C compiler for encrypted
computing.  The context consists of (i) a processor that `works
encrypted', taking in encrypted inputs and producing encrypted outputs
while the data remains in encrypted form throughout processing, and (ii)
machine codes that support arbitrary interpretations of the encrypted
input and outputs from each instruction, as far as an adversary who does
not know the encryption can tell.  The compiler on each recompilation of
the same source generates object code of the same form for which the
runtime traces have the same form, but the data beneath the encryption
may arbitrarily differ from nominal at each point in the trace,
independently so far as the laws of computation allow.
\vspace{-4ex}
}


\onecolumn \maketitle \normalsize \vfill

\section{\uppercase{Introduction}}
\label{s:Intro}

\noindent
This article describes practical `obfuscating' compil\-at\-ion from
ANSI C \cite{ansi99}
for {\em encrypt\-ed computing} \cite{BB13a}.  Encrypted co\-mputing
generates security via hard\-ware encryption for user data  
against {\em powerful insiders such as the
operator and operating system as adversaries}.\,\,Its
maj\-or component is a processor that `works encr\-ypt\-ed' in user mode
and works unencrypted in the operator mode in which operating system
code runs.  Enc\-ryption keys are installed at manufacture, as with
Smartcards
\cite{SmartCard}, or upload\-ed
in public view to a \mbox{write-only} internal
store via a Diffie-Hellman circuit \cite{buer2006cmos}, and are
not programmatically accessible. Theory in \cite{BB17a} shows
the operating system cannot directly or indirect\-ly by deterministic or
stochastic means read user data beneath the encryption, even
to a probability slightly above chance. That turns out to
imply it cannot be rewritten deliberately, even
stochas\-tically, to a value beneath the encryption that is
indep\-e\-n\-dently defined, such as $\pi$, or the encryption key.

Several fast
processors for encrypted computing are described in
\cite{BB18b}, including the authors' own
\KPU{} (Krypto Processor Unit) \cite{BB16b}, which
runs at appoximately the speed of a 433\,MHz classic Pentium,
embedding AES-128 \cite{DR2002}, and the slightly older
HEROIC \cite{heroic} which runs like a 25\,KHz Pentium, embedding
Paillier-2048 \cite{Pail99}.

But the machine code instruction set defining the programming interface
is also important.  A conventional
instruction set is insecure against powerful insiders who may steal an
(encrypted) user datum $x$ and put it through the machine's division
instruction to get $x/x$ encrypted, an encrypted 1.  Then any desired
encrypted $y$ may be constructed by repeatedly applying the machine's
addition instruction.  By using the instruction set's comparator
instructions (testing $2^{31}{\le}z$, $2^{30}{\le}z$, \dots) on an
encrypted $z$ and subtracting on branch, $z$ may be obtained
efficiently.  That is a {\em chosen instruction attack} (CIA)
\cite{Rass2016}. The instruction set has to 
resist that, but the compiler must be involved too, else
there would be {\em known
plaintext attacks} (KPAs) \cite{Biryukov2011} based on the idea that
not only do instructions like $x{-}x$ predictably
favor one value over others (the result
there is always $x{-}x{=}0$), but human programmers intrinsically use values
like 0,\,1 more often than others. The compiler's job is to
undo those statistics.

The compiler described here generates object code that varies from
compilation to compilation of the same source code but always looks the
same to an adversary, the difference lying in encrypted constants that
the adversary cannot read.  Runtime traces also `look the same,' with
the same instructions (modulo encrypted constants) in the same order,
the same jumps and branches, reading from and writing to the same
registers.  But data beneath the encryption varies arbitrarily and
independently from compilation to compilation at each point in the
trace, subject only to the proviso that a copy instruction preserves 
value, and the variation at the start and end of a loop is equal.

The compiler does that even for object code such as
the single instruction that adds two numbers, compiled from source code
$x=y+z$.  That gives the following conditions on the machine code instruction
design described in \cite{BB17a} and Box\,\ref{tb:ins}: instructions must
execute atomically (1), or recent attacks such as Meltdown
\cite{Lipp2018meltdown} and Spectre \cite{Kocher2018spectre} against
Intel might become feasible, must work with encrypted values (2) or an
adversary could read them, and must (3) be adjustable via
embedded encrypted constants to offset the values beneath the encryption
arbitrarily.
The condition (4) is for the security proofs in
\cite{BB17a}, and amounts to different padding or blinding factors
for encrypted program constants and runtime values.

\begin{frameenv}[tb]
\begin{flushleft}
\small
\refstepcounter{mybox}
\rm Box \arabic{mybox}:
\footnotesize
Instructions must \dots
\label{tb:ins}
\end{flushleft}
\medskip
\begin{minipage}{0.99\textwidth}
\small\sl
\begin{enumerate}
\item[{\rm(1)}] be a black box from the programming
interface with no observable intermediate states;
\item[{\rm(2)}] take encrypted inputs to encrypted outputs;
\item[{\rm(3)}] be adjustable via (encrypted) embedded constants 
to effect any given offset in inputs and outputs beneath the
encryption;
\item[{\rm(4)}] not permit ciphertext collisions between 
encrypted constants and runtime encrypted values.
\end{enumerate}
\end{minipage}
\setcounter{equation}{4}
\end{frameenv}

\customlabel{e:dagger}{$\dagger$}
\customlabel{e:star}{$\star$}
\customlabel{e:S}{\textsection}
\customlabel{e:ddagger}{$\ddagger$}

The effect of (1-4) is that an adversary not privy to the encryption
can feasibly believe nearly anything of the
data values beneath the encryption in a runtime trace (\eqref{e:star} of
Box\,\ref{tb:conj}).
By (3), an arbitrary variation from nominal introduced 
in one instruction could be corrected and changed again
in the next instruction, and (1-2) prevent the adversary 
noticing directly.  {\em The compiler's job} then boils down
to varying the encrypted program constants from recompilation to
recompilation so that all the feasible variations in the runtime data
are not only achieved in some
recompilation but also {\em equiprobable across recompilations}
(\eqref{e:S} of Box\,\ref{tb:conj}).  It is shown in \cite{BB17a}
that implies {\em cryptographic semantic security
{\rm\cite{Goldwasser1982}} for user data against insiders not privy to
the encryption} (\eqref{e:ddagger} of Box\,\ref{tb:conj}).
I.e., {\em encrypted
computation does not compromise the encryption}.

\begin{frameenv}[t]
\begin{flushleft}
\small
\refstepcounter{mybox}
\rm Box \arabic{mybox}:
\footnotesize
The\,role\,of\,the\,compiler.  Axioms\,(1-4)\,imply:
\label{tb:conj}
\end{flushleft}

\begin{minipage}{0.99\columnwidth}
\small\sl
\begin{enumerate}
\item[{\normalsize\eqref{e:star}}]
The same program code and runtime trace
can be
inter\-preted arbitrarily
with respect to the  plaintext data beneath the encryption at
any point in memory and in the control graph
by any observer and experimenter
who does not have the key to the encryption, with the proviso that
copy instructions preserve value and the variation from nominal
at start and end of a loop is the same.

{\rm With that, the compiler's job is to ensure:}

\item[{\normalsize\eqref{e:S}}]
All feasible plaintext interpretations of the plaintext data
beneath the encryption in
a program code and trace are possible and equiprobable.

{\rm Then an argument in \cite{BB17a} shows:}

\item[{\normalsize\eqref{e:ddagger}}]
Semantic security: no attack does better than guesswork
{\rm\cite{Goldwasser1982}} for insider adversaries not privy to
the encryption.
\end{enumerate}

\end{minipage}

\end{frameenv}

How the compiler achieves the `equiprobable variation' required by
\eqref{e:ddagger}  is encapsulated in Box~\ref{tb:how}.

It generates a new {\em obfuscation scheme} for the
object code each recompilation.
That is an {\em offset for the data beneath the
encryption in every memory and register location per every point in the
program control graph}.

Specifically, the compiler $\C{-}$ translates an expression $e$ that
is to end up in register $r$ at runtime into machine code $\it mc$
and generates a 32-bit offset $\Delta e$ for $r$ at the point in the
program where
it is loaded with the result of the expression $e$. That is

\begin{equation}
\C{e}^r = ({\it mc},\Delta e)
\end{equation}

The offset $\Delta e$ is the part of the obfuscation scheme relating to
register $r$ at the point where the encrypted value of
the expression is written to it.

Let $s(r)$ be the content of register $r$ in state $s$ of the processor
at runtime. The machine code {\em mc}'s action 
changes state $s_0$ to an $s_1$ with a ciphertext in $r$ whose plaintext
value differs by $\Delta e$ from the nominal value $e$ (bitwise
exclusive-or or the operator of another mathematical\,group are
alternatives to addition here):

\begin{equation}
s_0 \mathop\rightsquigarrow\limits^{\it mc}  s_1 ~\text{where}~ s_1(r) = 
\E[e + \Delta e]
\end{equation}

The encryption $\E$  is shared with the user and the processor (but
not the potential adversary, the operator and operating system). The
randomly generated offsets
$\Delta e$ of the obfuscation scheme are known to the user, but not
the processor and not the
operator and operating system. The user compiles the program and sends
it to the processor to be executed and needs to know the
offsets on the inputs and outputs.  That allows the right
inputs to be created and sent off for processing on the encrypted
computing platform, and allows sense to be made of the outputs received
back.

\subsection{Article Organisation}

Section~\ref{s:FxA} introduces a modified OpenRISC
(\url{http://openrisc.io}) machine code instruction set for encrypted
computing satisfying (1-4) first described in \cite{BB17a}, and its abstract
semantics.  That will be used instead of assembly language
to represent machine code, so readers need not face
assembler.

What is the difficulty solved here?  It is that nobody knows how to compile
satisfying (A-C) of Box\,\ref{tb:how}, or even if it is possible.
\cite{BB17a} showed how to do it for structured `if' and `while'
statements in C, assignments to variables, sequences of statements, and
function calls and return.  The only plaintext type treated was the
32-bit signed integer type (`int').  But extension to such C features
as pointers (memory addresses) is difficult, since an obfuscating
offset in a memory address amounts to a miss, where an offset in the
value in that location is correctable.

\begin{frameenv}[t]
\begin{flushleft}
\small
\refstepcounter{mybox}
\rm Box \arabic{mybox}:
\footnotesize
What the compiler does to achieve \eqref{e:S}:
\label{tb:how}
\end{flushleft}
\begin{enumerate}
\item[(A)] {\sl change only encrypted program constants, generating
via (3) an} {\rm offset from nominal values} {\sl for its inputs and outputs
beneath the encryption at each instruction, at each recompilation};
\end{enumerate}

\begin{enumerate}
\item[(B)] {\sl ensure runtime traces look unchanged, apart
from differences in program constants} (A);
\end{enumerate}

\begin{enumerate}
\item[(C)] {\sl generate\,all\,arrangements\,of\,offsets\,satisfying}\,(A), (B) {\sl eq\-ui\-prob\-ably. Each is an}
{\rm obfuscation scheme}.
\end{enumerate}

\end{frameenv}

Section~\ref{s:Comp} resumes obfuscating compilation
up to \cite{BB17a}, then extends it to nearly all ANSI C,
including pointers and arrays, union and `struct' (record) types,
floats, double length integers and floats, short integers and chars,
both signed and unsigned, gotos, and the interior functions extension.

\subsection{{Notation}}
\label{s:Not}

\noindent
Encryption is denoted by $x'=\E[x]$ of plaintext value $x$, sometimes
just $x'$. Decryption is $x=\D[x']$. The operation on the
ciphertext domain corresponding to $f$ on the plaintext domain is
written $[f]$, where $x'\mathop{[f]}y'=\E[x\mathop{f} y]$. Given
register $r$, $r^s$ is the next in sequence. Encrypted offsets
$\E[\Delta x]$\,may\,be\,compactly\,written $\Delta'_x$.

\vspace{-2ex}

\section{\uppercase{FxA Instructions}}
\label{s:FxA}

\noindent
A `fused anything and add' (FxA) 
\cite{BB17a} ISA is the compilation target here,  satisfying
conditions (1-4) of Section~\ref{s:Intro}.  Most of the integer
portion is shown in Table~\ref{tb:1}. It is adapted from the
OpenRISC instruction set v1.1 \url{https://openrisc.io/or1k.html}. That
has about 200 instructions (6-bit opcode plus variable minor
opcodes) separated into single and double precision
integer and floating point and
vector subsets and the instructions are all 32 bits long. Instructions access
up to three 32 general purpose registers (GPRs), defined via
5-bit specifier fields.  A register arithmetic instruction such as addition
$x{\leftarrow}y{+}z$ will specify three GPRs in which $x$, $y$,
$z$ reside in $3\times5=15$ bits,
while the operation `$+$' is specified in the 6-bit primary opcode
plus an 11-bit minor.  The minor may specify if addition is signed
or unsigned, half, single or double precision, integer or floating point, for
example. One operand may be supplied as a
(`immediate') constant in the instruction itself.

\begin{table}[tbp]
\caption{Integer portion of FxA machine code instruction set for
encrypted working --  abstract syntax and semantics.}
\label{tb:1}
{\centering
\scriptsize
\begin{tabular}{@{}l@{\,}c@{\kern1pt}c@{\kern1pt}l@{\kern1pt}l@{~}l@{}}
\em op.&
    \multicolumn{4}{@{}l}{\em fields} 

     & \multicolumn{1}{@{}l}{\em \hbox to 0.39in {mnem.\hfill} semantics \hfill}\\
\hline\\[-1ex]
add &$\rm r_0$
  &$\rm r_1$
    &$\rm r_2$
      &\kern0pt$k'$

        &\hbox to 0.39in {add} $r_0{\leftarrow}
           r_1\kern0.5pt\mathop{[+]} r_2\mathop{[+]} k'$\\
sub &$\rm r_0$
  &$\rm r_1$
    &$\rm r_2$
      &\kern0pt$k'$

        &\hbox to 0.39in {subtract} $r_0{\leftarrow}
           r_1\kern0.5pt\mathop{[-]} r_2\mathop{[+]} k'$\\
mul &$\rm r_0$
  &$\rm r_1$
    &$\rm r_2$
      &\kern0pt$k'_0\kern0pt k'_1\kern0pt k'_2$

       &\hbox to 0.39in {multiply}
       $r_0{\leftarrow}(r_1\kern0pt\mathop{[-]}k'_1\kern0pt)\mathop{[\,*\,]}(r_2\kern0pt\mathop{[-]}k'_2\kern0pt)\mathop{[+]}k'_0$\\
div &$\rm r_0$
  &$\rm r_1$
    &$\rm r_2$
      &\kern0pt$k'_0\kern0pt k'_1\kern0pt k'_2$

       &\hbox to 0.39in {divide} 
       $r_0{\leftarrow}(r_1\kern0pt\mathop{[-]}k'_1\kern0pt)\mathop{[\div]}(r_2\kern0pt\mathop{[-]}k'_2\kern0pt)\mathop{[+]}k'_0$\\
\dots\\
mov&$\rm r_0$&$\rm r_1$& &   

     &\hbox to 0.39in {move}
       $r_0{\leftarrow} r_1$\\
beq& $j$&$\rm r_1$&$\rm r_2$&$k'$  

     &\hbox to 0.39in {branch}
       ${\rm if}\,$b$\,{\rm then}\,{\it pc}{\leftarrow}{\it pc}{+}j$,
       $b\Leftrightarrow r_1 \mathop{[=]} r_2 \mathop{[+]} k'$\\
bne& $j$&$\rm r_1$&$\rm r_2$&$k'$  

     &\hbox to 0.39in {branch}
       ${\rm if}\,$b$\,{\rm then}\,{\it pc}{\leftarrow}{\it pc}{+}j$,
       $b\Leftrightarrow r_1 \mathop{[\ne]} r_2 \mathop{[+]} k'$\\
blt& $j$&$\rm r_1$&$\rm r_2$&$k'$   

     &\hbox to 0.39in {branch}
       ${\rm if}\,$b$\,{\rm then}\,{\it pc}{\leftarrow}{\it pc}{+}j$,
       $b\Leftrightarrow r_1 \mathop{[<]} r_2 \mathop{[+]} k'$\\
bgt& $j$&$\rm r_1$&$\rm r_2$&$k'$   

     &\hbox to 0.39in {branch}
       ${\rm if}\,$b$\,{\rm then}\,{\it pc}{\leftarrow}{\it pc}{+}j$,
       $b\Leftrightarrow r_1 \mathop{[>]} r_2 \mathop{[+]} k'$\\
ble& $j$&$\rm r_1$&$\rm r_2$&$k'$   

     &\hbox to 0.39in {branch}
       ${\rm if}\,$b$\,{\rm then}\,{\it pc}{\leftarrow}{\it pc}{+}j$,
       $b\Leftrightarrow r_1 \mathop{[\le]} r_2 \mathop{[+]} k'$\\
bge& $j$&$\rm r_1$&$\rm r_2$&$k'$   

     &\hbox to 0.39in {branch}
       ${\rm if}\,$b$\,{\rm then}\,{\it pc}{\leftarrow}{\it pc}{+}j$,
       $b\Leftrightarrow r_1 \mathop{[\ge]} r_2 \mathop{[+]} k'$\\
\dots \\
b   &$j$&   &   &   

     &\hbox to 0.39in {branch}
       ${\it pc}\leftarrow {\it pc}+j$\\
sw   &\multicolumn{4}{@{}l@{}}{$(k_0'){\rm r_0}~{\rm r_1}$}
     &\hbox to 0.39in {store}
       $\mbox{\rm mem}\llbracket r_0\mathop{[+]}k_0'\rrbracket \leftarrow r_1$ \\
lw   &\multicolumn{4}{@{}l@{}}{${\rm r_0}~(k_1'){\rm r_1}$}
     &\hbox to 0.39in {load}
       $r_0 \leftarrow \mbox{\rm mem}\llbracket r_1\mathop{[+]}k_1'\rrbracket
              $\\
jr   &$\rm r$&  &    &
     &\hbox to 0.39in {jump}
       ${\it pc} \leftarrow r$\\
jal  &$j$&      &    &
     &\hbox to 0.39in {jump}
       ${\it ra} \leftarrow {\it pc}+4;~{\it pc} \leftarrow j$\\
j   &$j$&      &    &
     &\hbox to 0.39in {jump}
       ${\it pc} \leftarrow j$\\
nop &   &      &    &
     &\hbox to 0.39in {no-op}\\[2ex]
\end{tabular}
}% centering

\noindent
\begin{minipage}{0.99\columnwidth}
{\scriptsize\sc Legend}\\[0.5ex]
\scriptsize
\noindent
\begin{tabular}{@{}l@{~}c@{~}l@{\quad}l@{~}c@{~}l@{\quad}l@{~}c@{~}l@{}}
r &--&register indices
 & $k$&--&32-bit integers
 & pc &--&prog.\ count\ reg.\\
$j$&--&program count or incr.
 & `$\leftarrow$'&--&assignment
 & ra &--&return addr.\ reg.\\
$\enc[\,~\,]$&--&encryption
 & $\dec[\,~\,]$&--&decryption
 & $r$ &--& register content \kern-5pt \\
$k'$&--&encrypted value $\enc[k]$
 &  \multicolumn{3}{@{}l@{}}{$x'\mathop{[g]}y' =
         \E[x\mathop{g}y]$}\kern-20pt
 &  \multicolumn{3}{@{}l@{}}{$x'\mathop{[R]}y' \Leftrightarrow
         x\mathop{R}y$}\kern-20pt
\end{tabular}
\end{minipage}
\end{table}

FxA instructions may contain 128-bit or more encrypted
constants, so some shoehorning is required.  A
`prefix' instruction takes care of that, supplying 
extra bits as necessary.  The prefix instruction is 32 bits long,
but several may be concatenated.

In addition to the integer instructions of Table~\ref{tb:1}, 
there are floating
point instructions {addf}, {subf}, {mulf} etc.\ paralleling the
OpenRISC floating point subset.  Contents of
registers and memory are encryptions of 32-bit integers that
encode floating point numbers (21 mantissa bits, 10 exponent
bits, 1 sign bit) via the IEEE\,754 standard encoding.
Let $*_f$ be the floating point multiplication on
plaintext integers encoding IEEE\,754 floats, and let
$[*_f]$ be the corresponding operation in the ciphertext domain,
following Section~\ref{s:Not}.  Then the floating
point multiplication instruction, has the semantics in
\eqref{e:dagger*f}:

\begin{equation}
\tag{$\dagger*_f$}
\label{e:dagger*f}
r_0 \leftarrow (r_1 \mathop{[-]} k'_1) \,\mathop{[*_f]}\, (r_2 \mathop{[-]} k'_2) \mathop{[+]} k'_0
\end{equation}

The $-$ and $+$ are the ordinary
plaintext integer subtraction and addition operations respectively,
and $[-]$ and $[+]$ are the corresponding operations in the
ciphertext domain, following Section~\ref{s:Not}. In words,
the FxA floating point multiplication takes the encrypted integers
representing floating point numbers and first offsets them as integers,
then multiplies them as floats, finally offsetting as integer again.
The operation is atomic, leaving no trace if aborted.

Our own FxA set has an improvement for runtime efficiency.
Floating point comparisons (indeed, also unsigned
integer comparisons) have two encrypted constants (the signed
integer comparisons in Table~\ref{tb:1} only have one).
The floating point branch-if-equal
instruction, for example, internally calculates \eqref{e:dagger=f}:

\begin{equation}
\tag{$\dagger{=}_f$}
\label{e:dagger=f}
(r_1\mathop{-}k_1') \,\mathop{[=_f]}\, (r_2\mathop{-} k_2')
\end{equation}

where $=_f$  is the floating point comparison on integers encoding
floats via IEEE\,754, and $[=_f]$ is the corresponding
test in the ciphertext domain. That is, $x'\mathop{[=_f]}y'
\Leftrightarrow x=_f y$. 

Condition (2) of Section~\ref{s:Intro} implies there should be
one more constant, an
encrypted bit $k_0'$ that decides if the 1-bit result of the
test is to be inverted. That is because the test outcome is observable
by whether the branch is taken, so by condition (3) of
Section~\ref{s:Intro} it should be varied by an encrypted
constant in the instruction.   Our own FxA instruction set does not
have that constant, because the variation is taken care of by the
compiler.  It
independently and randomly decides each time if the underlying source
code condition is to be interpreted as though by a {\em truthteller}, who
says `true' when true is meant and `false' when false is meant, or by a
{\em liar}, who says `false' when true is meant and `true' when false is
meant.  It equiprobably generates code for a lie and uses the
branch-if-not-equal instruction, or truthful code and uses
the branch-if-equal instruction.  The
compile procedure is described exactly in \cite{BB17a}.  With this
arrangement, whether the branch happens or not at runtime does not
even statistically relate to what the boolean result is meant to be.
Then condition (3) of Section~\ref{s:Intro} on the `output' of the
instruction being encrypted is satisfied vacuously, as there is
effectively no output.  An observer who sees a branch does not know if
that is the result of truthful interpretation of the source code and the
condition has come out true at runtime, or it is the result of
mendacious interpretation and the condition has come out false at
runtime.

A different FxA implementation may prefer to encode an extra bit in the
constants $k_1$, $k_2$, the encrypted versions $k_1'$, $k_2'$ of which
appear in the instruction, and use the exclusive or of the two to
determine if equals or not-equals is executed, satisfying (3) literally.

\section{\uppercase{Obfuscating Compilation}}
\label{s:Comp}

\noindent
The compiler works with a database $D : {\rm Loc}\,{\to}\,{\rm Int}$
containing the (32-bit) integer (type Int) offsets $\Delta l$ for data,
indexed per register or memory location $l$ (type Loc).  The offset
represents by how much the runtime data underneath the encryption is to
vary from nominal at that point in the program, and the database $D$
comprises the {\em obfuscation scheme}.  It is varied randomly by the
compiler as it makes its pass. The compiler
also maintains a database $L:{\rm Var}\to {\rm Loc}$ binding source
variables in registers and memory locations.

In \cite{BB17a}, a (non-side-effecting)
expression compiler that places its results in register $r$
is set out, of type:

\begin{align}
\S{\_ : \_}^r &:\hbox{DB}\times\hbox{Expr} \to \hbox{MC}\times\hbox{Int}
\end{align}

where MC is the type of machine code, a sequence of FxA instructions
{\em mc},
and Int contains the plaintext integer offset $\Delta$
 from nominal that the
compiler intends will exist in the result in $r$ beneath the encryption
when the machine code is evaluated at runtime. The aim is to satisfy
\eqref{e:S} by varying the offset $\Delta$ arbitrarily
and equiprobably.

To translate $x+y$, for example, where both $x$ and $y$ are signed integer
expressions, the compiler first 
emits machine code $\hbox{\em mc}_x^r$
computing expression $x$ in register $r$ with offset $\Delta x$. It then 
increments the register index $r$ to $r^s$ (the processor contains
registers allocated for this kind of temporary workspace calculation)
and emits machine code $\hbox{\em mc}_y^{r^s}$
computing expression $y$ in register $r^s$ with offset $\Delta y$. That
is

\begin{align*}
(\hbox{\em mc}_x^r,\Delta x) &= \S{D : x}^r\\
(\hbox{\em mc}_y^{r^s},\Delta x) &= \S{D : y}^{r^s}
\end{align*}
It then randomly decides an offset $\Delta$ for the whole expression
$x+y$ and emits the FxA integer addition instruction (the abstract
semantics $r \leftarrow r \mathop{[+]} r^s \mathop{[+]} k'$ is presented
here; notation as in Section~\ref{s:Not}) on the registers
$r$ and $r^s$ to return the result in $r$\/:

\begin{align}
\S{D: x+y}^r &= (\hbox{\em mc},\Delta)\\
\hbox{\em mc} &= \hbox{\em mc}_x^{r}; \hbox{\em mc}_y^{r^s};
                r \leftarrow r \mathop{[+]} r^s \mathop{[+]} k'\notag\\
           k  &= \Delta - \Delta x - \Delta y\notag
\end{align}

If the expression were instead $x*y$, then the emitted code terminates
with a multiplication instruction $c$:

\begin{align}
\S{D: x*y}^r &= (\hbox{\em mc},\Delta)
\label{e:C*}
\\
\hbox{\em mc} &= \hbox{\em mc}_x^{r}; \hbox{\em mc}_y^{r^s};c\notag\\
              c&=  r \leftarrow (r \mathop{[-]}\Delta'_x)
                             \mathop{[\,*\,]}
                             (r^s\mathop{[-]}\Delta'_y) \mathop{[+]}
                             \Delta'\notag
\end{align}

In every case the final offset $\Delta$ for the runtime result in $r$
beneath the encryption is freely generated \eqref{e:S}.

When an expression is a source code variable $z$ on its own, the lookup
database $L$ serves to locate where the variable is currently held
in registers and memory and the compiler will emit code to bring it
into register $r$ via an addition that introduces a new offset
$\Delta$, taking into account the offset $\Delta l = D l$ for the value in
storage in location $l = L z$.  A compiler abstraction layer called RALPH (Register
Abstraction Layer for Physical Hardware) optimises the
positioning of data and also provides an effectively unlimited supply of
temporary register space by aliasing-in use of the 64K extra `special
purpose registers' (SPRs) specified in the OpenRISC v1.1 architecture as
required. If those are exhausted, it will interpose use of
the execution stack (that is in memory).  The compilation of the subexpressions
$x$ and $y$ may use RALPH registers $r$, $r^s$, \dots and $r^s$,
$r^{ss}$,
\dots respectively, but not `lower' indices, so the code
generated for $x$ in $r$ is not interfered with  at runtime
by the code generated for $y$ in $r^s$.

The general rule is that the emitted code uses each
opportunity of a write instruction to generate a new offset $\Delta$ in
the written location, fulfilling \eqref{e:S}.

Statements are compiled differently to expressions because
they do not produce a result for which an offset from nominal is
to be generated. Instead they  have a side-effect.
Let Stat be the type
of statements, then compiling a statement 
returns not a single offset, as for expressions,
but a new obfuscation scheme:

\begin{align}
\S{\_ : \_} &:\hbox{DB}\times\hbox{Stat} \to \hbox{DB}\times\hbox{MC}
\end{align}

\noindent
Recall that the database $D$ of type DB holds the obfuscation
scheme (the offsets from nominal values in all locations, per
register or memory location).

Consider an assignment $x{=}e$ to a source code variable $x$, which the
location database $L$ says is bound in register $r{=}Lx$.
A pair in the cross product output will be written $D:m$ as syntactic
sugar.
First code $\hbox{\em
mc}_e$ for evaluating expression $e$ in temporary register {\bf t0} at
runtime is emitted via the expression compiler:
\[
(\hbox{\em mc}_e,\Delta e) = \S{D_0 : e}^{\bf t0}
\]
The offset $\Delta e$ is generated by the compiler 
for the result $e$ in {\bf t0}.  An instruction
to correct it to a new randomly chosen offset $\Delta$ in register $r$
is emitted next:

\begin{align}
\S{D_0: x{=}e}  &= D_1: ~\hbox{\em mc}_e ; r \leftarrow \hbox{\bf t0} \mathop{[+]} k'\\
                k &= \Delta - \Delta e\notag
\end{align}

The change made in the database of offsets $D_0$ to $D_1$ is at  the
entry for
location $r$, where the initial offset $D_0 r$ changes to $D_1 r = \Delta$,
the new offset being freely and randomly chosen by the compiler, supporting  \eqref{e:S}.

\subsection{{Other Base Types}}

\noindent
Double length (64-bit) `long long' integers are encod\-ed as two
32-bit integers.
The FxA instructions for dealing with them
contain (encrypted) 64-bit constants. Encryption
$x'=\E[x]$ is extended to 64-bit integers
as two encrypted 32-bit integers $x'=\E[x^H]{:}\E[x^L]$
where $x^H$, $x^L$
are the high and low 32 bits of the 64-bit plaintext integer $x$
respectively.

Let $-$ and $+$ be the
usual subtraction and addition  on 32-bit integers, and
let $-^2$ and $+^2$ be their application two-by-two to the pairs of 32-bit
integers comprising 64-bit integers. Let $*_{ll}$ be the
usual multiplication  on the 64-bit `long long' integers. 
The FxA 64-bit multiplication instruction has semantics:

\begin{equation}
\tag{$\dagger*_{ll}$}
x' \leftarrow \E[(y \,\mathop{-^2}\, k_1) \,\mathop{*_{ll}}\,
(z \,\mathop{-^2}\, k_2) \,\mathop{+^2}\, k_0]
\end{equation}

where $k_0$, $k_1$, $k_2$ are 64-bit plaintext integer constants with the
encrypted ciphertexts
$k'_i$, $i=0,1,2$ embedded in the instruction.

Putting it in terms of the effect on
register contents, the instruction semantics is:

\[
r_0^H{:}r_0^L \leftarrow (r_1^H{:}r_1^L \mathop{[-^2]}k_1')
                    \mathop{[*_{ll}]}
                  (r_2^H{:}r_2^L\mathop{[-^2]}k_2') \mathop{[+^2]}
                    k_0'
\]

However, for (encrypted) 64-bit operations the processor
partitions the register set into twos
referred to by one name each. In those terms the semantics
is:

\[
r_0 \leftarrow (r_1 \mathop{[-^2]}k_1')
                    \mathop{[*_{ll}]}
                  (r_2\mathop{[-^2]}k_2') \mathop{[+^2]}
                    k_0'
\]

When emitting the instruction, the compiler
understands that the registers $r$ listed are
really the firsts of pairs  $r$, $r^s$ containing the (encrypted) 64-bit integer
as two (encrypted) 32-bit halves, the high portion in $r$ and the low
portion in $r^s$ (bigendian order).
So for 64-bit multiplication 
the compiler emits not the instruction $c$ of \eqref{e:C*} in which
successive registers $r$ and $r^s$ are used for subexpressions $x$ and
$y$ respectively, but one in which the registers $r$ and $r^{ss}$
are used, respectively:

\begin{align}
c&=  r \leftarrow (r \mathop{[-^2]}\Delta'_ x)
                    \,\mathop{[*_{ll}]}\,
                  (r^{ss}\mathop{[-^2]}\Delta'_y) \mathop{[+^2]}
                    \Delta'
\label{e:C*l}
\end{align}

That leaves enough space for $x$ in $r$, $r^s$ and $y$ in $r^{ss}$,
$r^{sss}$.
The offset $\Delta$ is chosen freely, satisfying \eqref{e:S}.

As already set out for \eqref{e:dagger*f} and \eqref{e:dagger=f}, single
precision floats are encoded as 32-bit integers according to the
IEEE\,754 standard.  Double precision 64-bit floats (`double') are
encoded as two (encrypted) 32-bit integers, the top and bottom halves
respectively of a 64-bit integer following the IEEE\,754 coding for the
double. Let the $*_d$ operation be the double precision float
multiplication expressed on the 64-bit integer encodings of doubles as
two 32-bit integers, and let $[*_d]$ be the corresponding operation in
the cipherspace domain on two pairs of encrypted 32-bit integers. Then
as with double-length integer multiplication \eqref{e:C*l}, the
compiler emits not
instruction $c$ of \eqref{e:C*} with registers $r$ and $r^s$
for subexpressions $x$ and $y$ respectively, but

\begin{align}
c&=  r \leftarrow (r \mathop{[-^2]}\Delta'_x)
                    \,\mathop{[*_d]}\,
                  (r^{ss}\mathop{[-^2]}\Delta'_y) \mathop{[+^2]}
                    \Delta'
\end{align}

with $r$, $r^{ss}$, leaving room for double length
operands.

Machine code instructions that act on `short' (16-bit) or `char' (8-bit)
integers are not needed because short integers are promoted
to the standard 32-bit length at each operation in C.  The compiler has
a repertoire of translations for the source code cast
operation that change the 13 basic C types (signed/unsigned char, short,
int, long, long long integer, and float and double precision float, also
the single bit \_bool type) into one another.  A short is a (encrypted)
32-bit integer with the top 16 bits ignored. To cast a signed int to a
signed short, for example, the compiler in principle issues code that
moves the integer 16 places left and then 16 places right again:

\begin{align}
\S{D: (\hbox{short})e}^r  &= (\hbox{\em mc},\Delta)\\
                   (\hbox{\em mc}_e,\Delta e) &= \S{D: e}^r\notag\\
                   \hbox{\em mc} &= \hbox{\em mc}_e ;\notag\\
                            &\hbox{\quad} r \leftarrow (r \mathop{[-]} \Delta'_e)
                                 \mathop{[*]} \E[2^{16}] \mathop{[+]} k' ;\notag\\
                            &\hbox{\quad} r \leftarrow (r \mathop{[-]} k')
                                 \mathop{[\div]} \E[2^{16}]
                                 \mathop{[+]} \Delta' ;\notag
\end{align}

The constants $k$, $\Delta$ are freely chosen.

In order to avoid encryptions of numbers like $2^{16}$
appearing more than averagely often, the compiler in fact rather than emit the
literal sequence above causes a register $r^s$ to be loaded with the
encryption of
a random number $k_1$ and uses the instructions with $r^s\mathop{[-]}k_2'$
in place of
$\E[2^{16}]$, where $k_2{=}k_1{-}2^{16}$.  That makes the encrypted
constants that appear embedded in machine code unbiasedly
distributed (recall that (4) says they cannot serve as data in the
processor arithmetic).

For casts between integer and floating point types, the
instruction set provides atomic integer-to-float (and
vice versa) conversions.  In accord with the principle (3) of FxA
design, those embed encrypted constants that displace inputs
and outputs.

\subsection{{Arrays and Pointers}}

\noindent
It is possible to represent an array A as a set of individual variables
A0, A1, \dots and that allows the compiler to translate a lookup A[i] as
a compound expression `(i==0)?A0:(i==1)?A1:\dots', while  a write A[i]=x can
be translated to `if (i==0) then A0=x else if (i==1) then A1=x else
\dots'.  The entries get individual offsets from nominal $\Delta{\rm A0}$,
$\Delta{\rm A1}$, \dots in the obfuscation scheme maintained by the
compiler.

While that is logically correct, array access should be better than {\bf
O}n, so we have explored a different approach: entries of array A all share
the same offset $\Delta{\rm A}$ from their nominal value (beneath the
encryption).

Then pointer-based access becomes easier to generate code for. At compile time
where in the array the pointer will point is unknown, but 
the single shared offset may be used. The downside is that pointers p
must be declared with an array into which they point:

\begin{quote}
       restrict A int *p;
\end{quote}

With this approach, the compiler takes into account the offset for A when
constructing the dereference $*e$ of an expression $e$ that is a pointer
into A as follows.
It first emits code \hbox{\em mc}$_e$ that evaluates the
pointer with a randomly generated offset $\Delta e$ beneath the
encryption:

\begin{align*}
        (\hbox{\em mc}_e,\Delta e)&= \S{D: e}^r
\end{align*}

Then it emits a load instruction that compensates for the
offset $\Delta e$ in the address. 
The load instruction semantics is overall as follows -- but as the text below
explains, the\,runtime\,evaluation\,is\,not\,simple-minded:

\begin{align*}
    r &\leftarrow \hbox{mem}\llbracket r \mathop{[-]} \Delta'_e\rrbracket
\end{align*}

The processor at runtime has the
encrypted address $a'$ of the intended entry at address $a$ of
array\,A.

\begin{align*}
    a' &= r \mathop{[-]} \Delta'_e
\end{align*}

Despite encryption being one-to-many, in the case of the \KPU{} at
least, the processor's internal access
to keys makes that into a unique value $\hat a$ (not necessarily the
decrypted address $a$, though it is a unique hash of
that in the \KPU{}
in case of symmetric encryption; for Paillier it is a partial
decryption that removes an extra `blinding' multiplier from the ciphertext).

In the \KPU{}, the value $\hat a$ is memoised within a special front end to
the address
{\em translation lookaside buffer} (TLB) within the processor to a
value ${\rm TLB}(\hat a)$ selected at first encounter.
It was merely the 
next location free in a contiguous memory area in RAM, so there is no
mathematical relation of $a$ with ${\rm TLB}(\hat a)$, which
RAM gets as the address to look up. The load
instruction exact semantics is then as follows:

\begin{align*}
    r &\leftarrow \hbox{mem}[ {\rm TLB}(\hat a)]
\end{align*}

The TLB capacity in the \KPU{} is 
2M\,addresses. It is backed by an encrypted database in RAM,
and a missing memoisation causes 
a minor memory fault signal, the handler for which recovers the
encrypted information from RAM to the TLB.

The memoisation is
changed at every write to it, so RAM sees a random pattern of accesses
overall, but here for read
the existing memoisation ${\rm TLB}(\hat a)$
within the TLB is
used. The upshot is that the simple load instruction issued by the
compiler both works at runtime
despite RAM getting a dynamically varying address to lookup, and the
location containing
encrypted data, and also does not expose unencrypted information.

The value retrieved by the load instruction has the offset
$\Delta{\rm A}$ and the compiler emits an add instruction
to change it to a freely chosen offset $\Delta$, as follows:

\begin{align}
                       r &\leftarrow r \mathop{[+]} k'\notag
\end{align}

where~ $k  = \Delta - \Delta{\rm A}$.
The complete code emitted is:

\begin{align}
\S{D: *e}^r &= (\hbox{\em mc},\Delta)
        \label{e:pointer}\\
        \hbox{\em mc} &= \hbox{\em mc}_e;\notag\\
                      &\quad r \leftarrow \hbox{mem}\llbracket r
                           \mathop{[-]} \Delta'_e\rrbracket ;
                      \notag
                      &&\mbox{\# lw r $(\E[-\Delta_e])$r}
                      \\
                      &\quad r \leftarrow r\mathop{[+]}k';
                      &&\mbox{\# addi r r $k'$}
                      \notag
\end{align}
where $k{=}\Delta{-}\Delta{\rm A}$.

Then an indexed array lookup A[i] is handled by dereferencing a pointer *(A+i).
However, writing an array entry is more problematic, not because the
code is complicated (the final addition instruction in \eqref{e:pointer}
is replaced with an addition instruction involving the register $r^s$ in
which the overwriting value has been written), but because it
should on principle be coincident with a change of the offset $\Delta$A
for the target entry in the array, and therefore for every entry in
the array.  That means every array entry must be rewritten to the new offset
whenever one is written, an {\bf O}n  `write storm.' We are evaluating
the real costs -- 
only write bandwidth is occupied, not write latency increased,
since writes are asynchronous through the writeback cache. The
behaviour is like oblivious RAM (ORAM) \cite{ostrovsky1990},
a modified RAM that encrypts data and addressing, hiding programmed
accesses among random false accesses.

\subsection{{Structs}}

\noindent
C `structs' are records with fixed fields.  The approach the compiler
takes is to maintain a different offset per field, per variable of
struct type.  That is, for a variable x of struct type with fields a and b
the compiler maintains offsets $\Delta$x.a and $\Delta$x.b.
It is as though there were two variables, x.a and x.b respectively.

In the case of an array A the entries of which are structs
with fields a and b, the compiler maintains two separate offsets
$\Delta$A.a and $\Delta$A.b for the two fields of its entries, and so
on recursively if those fields are themselves structs. Updating one
field in one entry changes the offset and is accompanied by a `write
storm' of adjustments over the stripe through the array consisting of
that same field in all entries.  That is more efficient than a storm to
a whole array, so for more efficient computing in this context, array
entries should be split into structs whenever possible.

\subsection{{Unions}}

The obfuscation scheme in a union type such as
\[\tt
union~\{ struct \{int~a; float~b[2];\};
        double~c[2];
\}
\]
has to engage compatible obfuscation schemes for the component types.
The obfuscation scheme for the struct will have the pattern (in 32-bit
words) $x, y, y$, with $x$ the offset for the int and $y$ the offset
for the float array entries, while the pattern for the double
array will be $u, v, u, v$, repeating $u, v$ for each entry.
\begin{align*}
\tt union  \{&\tt struct
\{\underbracket{~\raisebox{0pt}[2ex][0.6ex]{$\tt
int~a$}}_{\raisebox{-1pt}[2ex][0pt]{$x$}}; 
                       \underbracket{\tt float~b[2]}_{\mbox{$y,y$}};\};
             \underbracket{\tt double~c[2]}_{\raisebox{-1pt}{$u,v,u,v$}}; \}
\end{align*}
The resolution is $x{=}y{=}u{=}v$ for a scheme $y,y,y,y$. That is
the least restrictive obfuscation scheme forced by the union
layout here.

\subsection{{Go To and Come To}}
\label{s:Goto}

\noindent
Gotos are treated in three steps: (i)
local labels x are declared as follows:

\begin{quote}
 \_\_label\_\_~x
\end{quote}

Variables y that might be affected (are read or written) after the
labelled point `x:\dots' in the program) by a goto~x must be
declared prior to the \_\_label\_\_~x declaration, because  the compiler
snapshots the obfuscation scheme (the offsets $\Delta$y) for the
variables y that are in scope at that point and binds it to x. Then
(ii) at each goto~x statement the compiler emits machine code add
instructions that change the offsets $\Delta$y back to the bound
scheme.  Further (iii), just before label x on the default `fall
through' control path, the compiler emits machine code add
instructions that changes the offsets $\Delta$y of the variables y back
to the scheme bound to x.  So a label is associated with a non-trivial
sequence of machine code, and it may be thought of as a source-level
`come to x' statement.

The reason for this approach is that converging control paths must agree
on the obfuscation scheme at the join.  At the point labelled by x,
paths via incoming gotos converge with the fall-through.  What
obfuscation scheme should be used there?  The one at the goto~x
or the one at the labelled point?  Either order would sometimes force
two passes.  The simplest approach is the one adopted: declare
the label and bind the obfuscation scheme in use at the declaration.

\subsection{{Interior Functions}}
\label{s:Interior}

\noindent
Interior functions are a common extension to ANSI C. They are
functions f declared within another function g. They
have access to g's arguments and local variables x.
At the point of declaration of f the compiler knows the obfuscation scheme
$\Delta$x in use for g's variables x and can emit code
for the interior function f that is accommodated to it.  Unfortunately,
g may call f at a point where the obfuscation
scheme for x is different, having been altered by the compiler at
some intervening write to x in g.

The same solution as for gotos works. The interior function 
declaration is treated as a label declaration:

\begin{quote}
 \_\_label\_\_~f
\end{quote}

The obfuscation scheme at that point is bound to f, and every call of
f in g is treated as though it were preceded by the `come
to f' statement of Section\,\ref{s:Goto}.  That reestablishes the
obfuscation scheme bound to f at its declaration.  A return from f
is accompanied by instructions that establish another obfuscation
scheme for the external's variables x.

\vspace{-1ex}

\section{\uppercase{Summary}}

\noindent
Every machine code write instruction emitted by the compiler
establishes a new offset $\Delta$ from the nominal value of the
plaintext data beneath the encryption in the target location at that
point in the program.  The $\Delta$s are randomly chosen by the
compiler, subject to the constraint that the offsets at the ends of
control paths agree at the join \eqref{e:S}.
A function call to f does establish the same obfuscation scheme after
every call, but different instances f$_i$ of f may be
compiled per call point $i$, each with separate final obfuscation schemes.

At the current stage of development, the compiler has near total
coverage of ANSI C and GNU extensions, including
statements-as-expressions and expressions-as-statements.  
Efficient strings (currently arrays of chars) are still to come.  It is
being debugged via the {\em gcc\/} `c-torture' testsuite v2.3.3, and we
are presently about one quarter through that.

\vspace{-1ex}

\section{\uppercase{Conclusion}}

\noindent
How to compile C in an `obfuscating' way for enc\-rypted
computing has
been set out.  That class of platform natively works on encrypted data
in user mode, never decrypting it as far as can be told via the
programming interface.  With a modified RISC instruction set for
encrypted computing, arbitrary assignments for the plaintext data beneath
the encryption in a single execution are feasible.  The obfuscating
compiler makes them all equiprobable.  Without it, stochastic
arguments are unavailable.  With it, semantic security (no attack does
better than guesswork) for user data beneath the encryption holds
by definition, given the encryption is independently secure.

\section*{\uppercase{Acknowledgements}}

\noindent
Peter Breuer wishes to thank Hecusys LLC ({hecusys.com}) for
continued support in \KPU{} development.

\vspace{-1ex}

\renewcommand{\baselinestretch}{0.93}
\bibliographystyle{apalike}
\begin{small}
\bibliography{secrypt2018b}

\begin{thebibliography}{}

\bibitem[Biryukov, 2011]{Biryukov2011}
Biryukov, A. (2011).
\newblock Known plaintext attack.
\newblock In van Tilborg, H. C.~A. and Jajodia, S., editors, {\em Encyclopedia
  of Cryptography and Security}, pages 704--705. Springer, Boston, MA.

\bibitem[Breuer and Bowen, 2013]{BB13a}
Breuer, P. and Bowen, J. (2013).
\newblock A fully homomorphic crypto-processor design: Correctness of a secret
  computer.
\newblock In {\em Proc.\ Int.\ Symp.\ Eng.\ Sec.\ Softw. Sys. ({ESSoS}'13)},
  number 7781 in LNCS, pages 123--138, Heidelberg/Berlin. Springer.

\bibitem[Breuer et~al., 2016]{BB16b}
Breuer, P., Bowen, J., Palomar, E., and Liu, Z. (2016).
\newblock A practical encrypted microprocessor.
\newblock In {\em Proc. 13th Int.\ Conf.\ Sec. and Cryptog. ({SECRYPT}'16)},
  volume~4, pages 239--250, Portugal. {SCITEPRESS}.

\bibitem[Breuer et~al., 2017]{BB17a}
Breuer, P., Bowen, J., Palomar, E., and Liu, Z. (2017).
\newblock On obfuscating compilation for encrypted computing.
\newblock In {\em Proc. 14th Int.\ Conf.\ Sec. and Cryptog. ({SECRYPT}'17)},
  pages 247--254, Portugal. {SCITEPRESS}.

\bibitem[Breuer et~al., 2018]{BB18b}
Breuer, P., Bowen, J., Palomar, E., and Liu, Z. (2018).
\newblock Superscalar encrypted {RISC}: The measure of a secret computer.
\newblock In {\em Proc.\ 17th Int.\ Conf.\ Trust, Sec.\ \& Priv.\ in Comp.\ \&
  Comms.\ (TrustCom'18)}. {IEEE} Comp.\ Soc.
\newblock To appear.

\bibitem[Buer, 2006]{buer2006cmos}
Buer, M. (2006).
\newblock {CMOS}-based stateless hardware security module.
\newblock {US}\,Pat.\ App.\ 11/159,669.

\bibitem[Daemen and Rijmen, 2002]{DR2002}
Daemen, J. and Rijmen, V. (2002).
\newblock {\em The Design of {R}ijndael: {AES} -- The Advanced Encryption
  Standard}.
\newblock Springer.

\bibitem[Goldwasser and Micali, 1982]{Goldwasser1982}
Goldwasser, S. and Micali, S. (1982).
\newblock Probabilistic encryption \& how to play mental poker keeping secret
  all partial information.
\newblock In {\em Proc.\ 14th Ann.\ {ACM} Symp.\ Th.\ Comp.}, {STOC}'82, pages
  365--377. ACM.

\bibitem[{ISO/IEC}, 2011]{ansi99}
{ISO/IEC} (2011).
\newblock Programming languages -- {C}.
\newblock 9899:201x Tech.\ Report n1570, International Organization for
  Standardization.
\newblock {JTC 1, SC 22, WG 14}.

\bibitem[Kocher et~al., 2018]{Kocher2018spectre}
Kocher, P., Genkin, D., Gruss, D., Haas, W., Hamburg, M., Lipp, M., Mangard,
  S., Prescher, T., Schwarz, M., and Yarom, Y. (2018).
\newblock Spectre attacks: Exploiting speculative execution.
\newblock {\em ArXiv e-prints}.

\bibitem[K\"ommerling and Kuhn, 1999]{SmartCard}
K\"ommerling, O. and Kuhn, M.~G. (1999).
\newblock Design principles for tamper-resistant smartcard processors.
\newblock In {\em Proc. {USENIX} Work.\ Smartcard Tech.}, pages 9--20.

\bibitem[Lipp et~al., 2018]{Lipp2018meltdown}
Lipp, M., Schwarz, M., Gruss, D., Prescher, T., Haas, W., Mangard, S., Kocher,
  P., Genkin, D., Yarom, Y., and Hamburg, M. (2018).
\newblock Meltdown.
\newblock {\em ArXiv e-prints}.

\bibitem[Ostrovsky, 1990]{ostrovsky1990}
Ostrovsky, R. (1990).
\newblock Efficient computation on oblivious {RAM}s.
\newblock In {\em Proc. 22nd Ann. {ACM} Symp.\ Th.\ Comp.}, pages 514--523.
  ACM, ACM.

\bibitem[Paillier, 1999]{Pail99}
Paillier, P. (1999).
\newblock Public-key cryptosystems based on composite degree residuosity
  clas\-ses.
\newblock In Stern, J., editor, {\em Proc.\ Int.\ Conf.\ Th.\ Appl. Cryptog.
  Techn. ({EUROCRYPT}'99)}, number 1592 in LNCS, pages 223--238,
  Heidelberg/Berlin. Springer.

\bibitem[Rass and Schartner, 2016]{Rass2016}
Rass, S. and Schartner, P. (2016).
\newblock On the security of a universal crypto\-computer: The chosen
  instruction attack.
\newblock {\em {IEEE} {A}ccess}, 4:7874--7882.

\bibitem[Tsoutsos and Maniatakos, 2015]{heroic}
Tsoutsos, N.~G. and Maniatakos, M. (2015).
\newblock The {HEROIC} framework: Encrypted computation without shared keys.
\newblock {\em {IEEE} Trans. {CAD} {IC} Sys.}, 34(6):875--888.

\end{thebibliography}
\end{small}
\renewcommand{\baselinestretch}{1}

\end{document}